\documentclass[conference]{IEEEtran}
\IEEEoverridecommandlockouts

\usepackage{amsmath,amssymb,amsfonts}
\usepackage{algorithmicx}
\usepackage{graphicx}
\usepackage{textcomp}
\usepackage{xcolor}
\usepackage[hidelinks]{hyperref}
\usepackage{algorithm}
\usepackage{algpseudocode}
\usepackage{float}
\usepackage{placeins}
\usepackage{enumitem}
\usepackage{bm}
\usepackage{cite}
\usepackage{booktabs}
\usepackage{float}
\usepackage{booktabs,tabularx,makecell}
\usepackage{caption}

\def\BibTeX{{\rm B\kern-.05em{\sc i\kern-.025em b}\kern-.08em
    T\kern-.1667em\lower.7ex\hbox{E}\kern-.125emX}}
\begin{document}

\title{UAV Trajectory Optimization via Improved Noisy Deep Q-Network\\
}
\author{
\IEEEauthorblockN{1\textsuperscript{st} Zhang Hengyu}
\IEEEauthorblockA{\textit{Dept. of Computer Science} \\
\textit{Wenzhou-Kean University}\\
Wenzhou, China \\
zhanheng@kean.edu}
\and
\IEEEauthorblockN{2\textsuperscript{nd} Maryam Cheraghy*}
\IEEEauthorblockA{\textit{Dept. of Computer Science} \\
\textit{Wenzhou-Kean University}\\
Wenzhou, China \\
mcheragh@kean.edu \\
*Corresponding author}
\and
\IEEEauthorblockN{3\textsuperscript{rd} Liu Wei}
\IEEEauthorblockA{\textit{Dept. of Computer Science} \\
\textit{Wenzhou-Kean University}\\
Wenzhou, China \\
lwei@kean.edu}
\and
\IEEEauthorblockN{4\textsuperscript{th} Armin Farhadi}
\IEEEauthorblockA{\textit{School of Electrical and Computer Engineering} \\
\textit{University of Tehran}\\
Tehran, Iran \\
armin.farhadi@ut.ac.ir}
\and
\IEEEauthorblockN{5\textsuperscript{th} Meysam Soltanpour}
\IEEEauthorblockA{\textit{School of Computer Science and Info. Eng.} \\
\textit{Qilu Institute of Technology}\\
Jinan, China \\
soltanpourmeysam1@gmail.com}
\and
\IEEEauthorblockN{6\textsuperscript{th} Zhong Zhuoqing}
\IEEEauthorblockA{\textit{Dept. of Computer Science} \\
\textit{Wenzhou-Kean University}\\
Wenzhou, China \\
zhongzh@kean.edu}
}

\makeatletter
\renewcommand{\IEEEaftertitletext}{\vspace{-1.5em}} 
\makeatother

\maketitle
\begin{abstract}
This paper proposes an Improved Noisy Deep Q-Network (Noisy DQN) to enhance the exploration and stability of Unmanned Aerial Vehicle (UAV) when applying deep reinforcement learning in simulated environments. This method enhances the exploration ability by combining the residual NoisyLinear layer with an adaptive noise scheduling mechanism, while improving training stability through smooth loss and soft target network updates. Experiments show that the proposed model achieves faster convergence and up to $+40$ higher rewards compared to standard DQN and quickly reach to the minimum number of steps required for the task — $28$ in the $15 \times 15$ grid navigation environment set up. The results show that our comprehensive improvements to the network structure of NoisyNet, exploration control, and training stability contribute to enhancing the efficiency and reliability of deep Q-learning.
\end{abstract}
    

\section{Introduction}
Unmanned aerial vehicles (UAVs) have become a key technology in fields such as environmental monitoring and emergency rescue \cite{Almahamid2022}, and can significantly enhance communication coverage in massively connected 3D wireless networks~\cite{10105221,10453236}. As the application of UAV expands to more complex and dynamic environments, the demand for reliable autonomous navigation systems has significantly increased \cite{Almahamid2022}. These systems not only need to ensure the effective path planning of UAVs, but also need to guarantee their strong communication, real-time decision-making, and adaptability to environmental uncertainties. However, achieving these goals in environments with limited signals is particularly challenging, such as in some densely populated urban areas, disaster-stricken regions, or remote areas. Because traditional navigation algorithms struggle in the face of limited sensing, sparse rewards, and unstable policies \cite{Wang2020}. Therefore, developing intelligent UAV navigation methods that can optimize flight trajectories while ensuring communication quality has become an important research direction.

So far, numerous research efforts on reinforcement-learning-based UAV navigation have been reported in the literature. For example, Double Deep Q-Network (DQN) was introduced in \cite{Yang2020} to mitigate Q-value over-estimation, while Prioritized Experience Replay (PER) was employed in~\cite{Hu2022} to accelerate convergence in cluttered indoor scenes. NoisyNet layers were incorporated into DQN in~\cite{Yang2025} to enrich exploration without manual $\epsilon$-schedules, whereas a distributed privileged Reinforcement Learning (RL) framework leveraged high-fidelity states for the critic only during training, enabling zero-shot transfer to GPS-denied mazes in~\cite{Wang2024Privileged}. The influence of reward sparsity and obstacle density on path length and collision rate was systematically evaluated in~\cite{Wang2024OJVT} for single-UAV missions, and multi-UAV cooperation under limited bandwidth was investigated via attention-based Counterfactual Multi-Agent Policy Gradients (COMA) in~\cite{Zhao2025}. However, none of~\cite{Yang2020,Hu2022,Yang2025,Wang2024Privileged,Wang2024OJVT, Zhao2025} considers adaptive noise scheduling together with residual NoisyLinear layers under realistic communication-loss constraints. Performance analyses that jointly account for stochastic packet drops and heterogeneous signal zones are scarce, being limited to the hexagonal-cell study of~\cite{Cho2022} and the ideal-vision assumption of~\cite{Ou2020}. RL-driven computation offloading frameworks such as~\cite{muslim2022reinforcement} also highlight the importance of jointly optimizing communication reliability and decision-making under heterogeneous edge–cloud conditions. Therefore, this paper proposes an improved Noisy DQN scheme that couples residual NoisyLinear blocks with self-tuning Gaussian noise and Double DQN. The main contributions of this work are as follows:
\begin{itemize}
    \item A communication-constrained UAV navigation problem is formulated on a $15{\times}15$ grid, and a reward is constructed to jointly encode goal proximity, collision margins, and minimum signal strength under distance- and obstacle-induced attenuation.
    \item An improved Noisy DQN architecture is presented, where residual NoisyLinear blocks with factorized Gaussian noise are coupled with an adaptive, performance-aware noise schedule and periodic re-sampling, yielding state-dependent and trainable exploration.
    \item Training stability is strengthened by combining Double-DQN target estimation and soft target updates, by which overestimation and target drift are mitigated.
    \item A comprehensive evaluation is conducted on the designed environment, where faster convergence, higher rewards, and fewer steps than a DQN algorithm and their variants are consistently observed.
\end{itemize}
The remainder of this paper is organized as follows. Section~II presents the system model. Section~III states the problem formulation. Section~IV details the proposed Improved Noisy DQN method and the complete training algorithm. Section~V reports the simulation setup and comparative results. Section~VI offers conclusions and directions for future work.
\section{System Model}
We model the UAV trajectory planning problem under communication constraints as an optimization task within a two-dimensional grid-based environment. The objective is to find a feasible trajectory that minimizes travel cost while maintaining reliable signal quality and avoiding obstacles. To facilitate simulation, the environment is defined as an \(A \times A\) discrete grid space, resulting in a total of \(A^{2}\) grid cells. The UAV starts from a fixed starting point \((0,0)\), and the goal is to autonomously plan the path within the maximum number of steps \(T_{\max}\) to reach the predetermined target position \((A-1, A-1)\). Throughout the paper, we use symbols \(A\) and \(T_{\max}\) in the formulation (e.g., constraints) and provide their numerical values only in the \emph{Simulation Results} section. We refer to “grid cells” (or “discrete positions”), since the grid is fixed and the UAV moves.
\begin{figure}[htbp]
    \centering
    \includegraphics[width=0.30\textwidth]{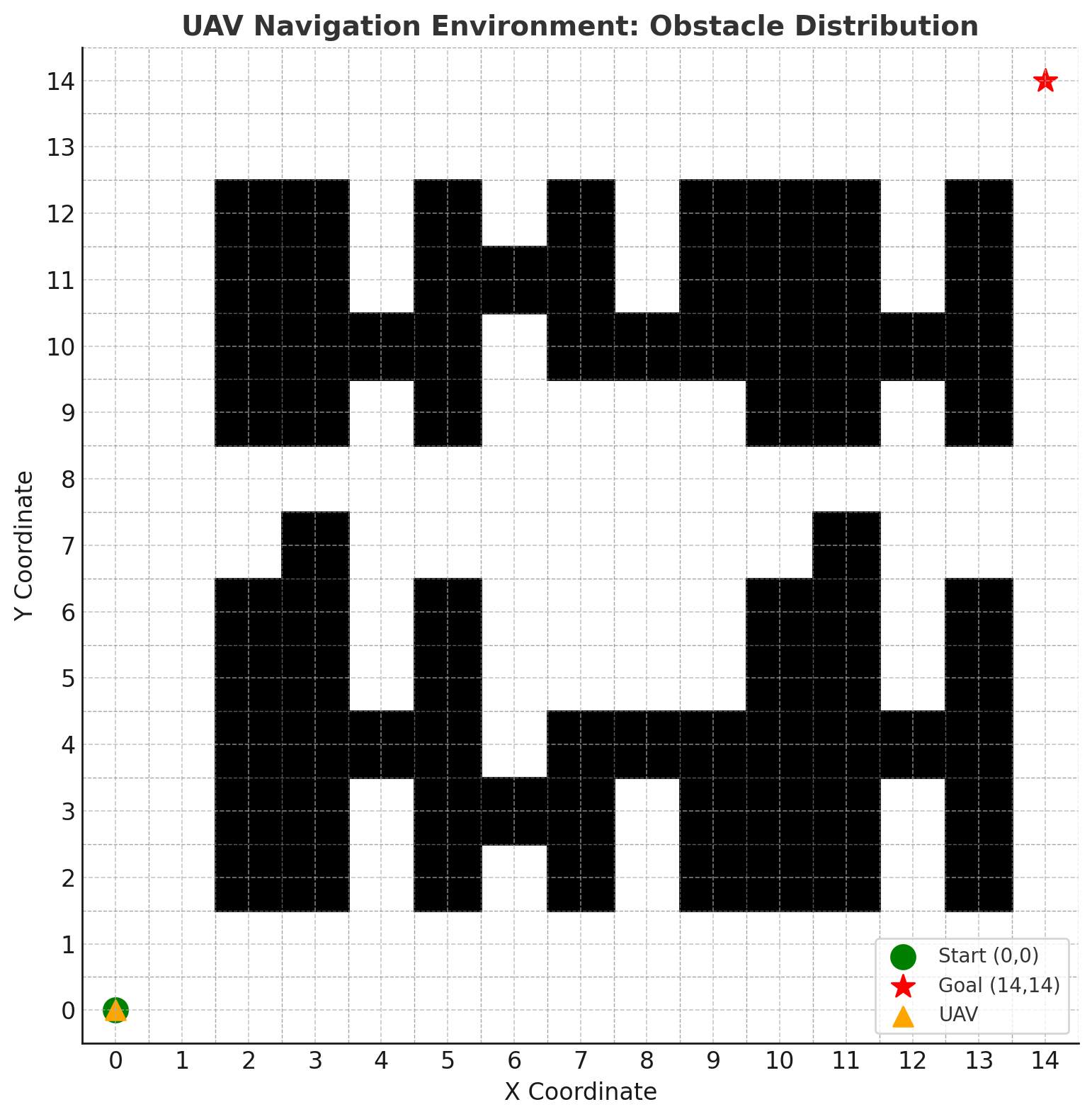} 
    \caption{Obstacle distribution map in the $15 \times 15$ UAV navigation environment.}
    \label{fig:obstacle_map}
\end{figure}
To increase the realism and difficulty of the navigation task, obstacles are distributed throughout the grid in a structured yet diverse manner. The corresponding obstacle layout is visualized in Fig.~\ref{fig:obstacle_map}, providing a structured yet flexible navigation space for the UAV. Specifically, vertical obstacle corridors are placed along the left, central, and right regions of the map, each with intentional gaps to preserve multiple feasible paths from the start to the goal. Additional horizontal obstacles are strategically inserted to further complicate route selection and prevent trivial solutions. Randomly positioned obstacles are also added to introduce variability between episodes. Importantly, both the start and goal areas are kept free of obstacles to ensure that the UAV can always begin and complete its mission. This design creates a challenging environment that requires the agent to explore, adapt, and optimize its trajectory in the presence of multiple potential routes and signal constraints.\\
To more realistically reflect the challenges of UAV deployment in communication-constrained environments, the simulation incorporates a signal attenuation mechanism in which signal strength dynamically decays with distance from the base station and is further degraded by surrounding obstacles. This model introduces an implicit constraint into the navigation task, requiring the agent to maintain reliable communication while optimizing its path. 
\vspace{-4pt}
 \section{Problem Formulation}
We formulate a discrete optimization problem on an $A\times A$ grid: the decision variables are $\{(x_n,y_n)\}_{n=0}^{A-1}$; the objective is to reach $(A-1,A-1)$ with the fewest steps (equivalently, maximum return) subject to collision-avoidance and minimum signal-strength constraints, as detailed in \eqref{eq:opt_problem}. Since the position of the UAV is step-dependent, the optimization problem can be considered as a trajectory optimization for the UAV. Hence, below we construct an optimization problem:
\begin{equation}
\begin{aligned}
    \min_{x_n,y_n} \; & D_n \\
    \max_{x_n,y_n} \; & \left\| (x_n,y_n)^T - (x_{n-1},y_{n-1})^T \right\|^2 \\
    \text{s.t.} \quad 
    \text{C1: } \; & \left\| (x_n,y_n)^T - (x_{n-1},y_{n-1})^T \right\|^2 \leq T_{\max}, \\
    \text{C2: } \; & (x_n,y_n)^T \neq (x_i,y_i)^T, \quad \forall i \in \mathcal{I}, \\
    \text{C3: } \; & x_n,  y_n \in \{0, 1, ..., A-1\}, \\
    \text{C4: } \; & \left\| (x_n,y_n)^T - (x_{i},y_{i})^T \right\|^2 \geq D_{\min}, \\
    \text{C5: } \; & \sigma_n\ > \sigma_{\min},
\end{aligned}
\label{eq:opt_problem}
\end{equation}
where $(x_n, y_n)^T$ represents the current UAV position, $(x_{i}, y_{i})^{T} $ denotes the obstacles position. In the optimization problem \eqref{eq:opt_problem}, $D_n$ is the distance between the goal and the current position of the UAV, defined as follows:
\begin{align}
   & D_n = \left\| (x_n,y_n)^T - (x_{goal},y_{goal})^T\right\|^2.
\label{D_n}
\end{align}
Moreover, $\mathcal{I}$ denotes the set of obstacles available in the investigated environment, which can be expressed as $\mathcal{I} = \{1, 2, \dots, I\}$. Finally, $D_{\min}$ represents the minimum distance between the UAV and the obstacles. In this paper, we set it to $2.5$ (i.e., $D_{\min} = 2.5$), and we model the signal quality as a smooth, isotropic Gaussian proxy centered around the base station. It decays with the square of the distance from the base station.
{
\setlength{\abovedisplayskip}{3pt}
\setlength{\belowdisplayskip}{3pt}
\setlength{\abovedisplayshortskip}{3pt}
\setlength{\belowdisplayshortskip}{3pt}
\begin{equation}
\sigma_n = e^{-\beta\,\big\|(x_n,y_n)^{\top}\big\|^{2}}, \qquad \beta=2.5 .
\label{eq:sigma}
\end{equation}
}
The investigated problem \eqref{eq:opt_problem} is a multi-objective optimization problem. Because the optimization problem has two objective functions \cite{farhadi2024meta}. In this problem, we aim to reach the goal within a limited number of steps while considering environmental constraints such as avoiding collisions with obstacles. In the investigated problem, $\text{C1}$ indicates that the maximum location change compared to the previous step must not exceed $T_{\max}$. $\text{C2}$ guarantees that the current UAV position cannot coincide with the position of obstacles. $\text{C3}$ defines the minimum and maximum allowable values for $(x_n,y_n)^T$, and these values are discrete as mentioned in the constraint. Since the signal attenuation increases when the UAV approaches the obstacles, we introduce a constraint regarding this in $\text{C4}$. Finally, the signal strength must be greater than $\sigma_{\min}$, as specified in $\text{C5}$. The optimization problem \eqref{eq:opt_problem} is a non-convex multi-objective optimization problem due to its non-convex quadratic constraints. To solve this problem using conventional optimization methods, we would need to convert it into a convex problem. This conversion is challenging, since it requires finding an appropriate method, algorithm, approximation, or variable transformation~\cite{10778205,10436671}. Once converted, the problem could then be solved with classical convex optimization techniques. However, our problem involves multiple steps to reach the optimal position. For this reason, it is preferable to use RL methods to find a proper trajectory for the UAV without needing to convert the problem into a convex form.  
It is also worth noting that convex programming is very time-consuming and not suitable for environments that change dynamically~\cite{10778205}. This limitation provides another motivation for using RL to solve this problem.\\
\begin{table}[!t]
\vspace{-0.75cm}
\caption{Base Variables}
\vspace{-4pt}
\label{tab:base_vars_core_strict}
\footnotesize
\centering
\renewcommand{\arraystretch}{0.85}
\setlength{\extrarowheight}{0pt}
\begin{tabularx}{\columnwidth}{l >{\raggedright\arraybackslash}p{0.45\columnwidth} >{\raggedright\arraybackslash}X}
\toprule
\textbf{Symbol} & \textbf{Meaning} & \textbf{Value/Range} \\
\midrule
$A$ & Grid size (square map with $A\times A$ cells) & 15 \\
$T_{\max}$ & Episode step limit & 40 \\
$D_{\min}$ & Minimum allowed squared distance to any obstacle & 2.5 \\
$(x_n, y_n)$ & The position of the UAV in the grid & $\{(x_n, y_n)\}_{n=0}^{A-1}$ \\
$(x_{goal}, y_{goal})$ & The target position of the UAV in the grid & (14,14)) \\

\midrule
$\sigma_n$ & Signal-strength proxy at $n$ & $e^{-2.5\,\big\|(x_n,y_n)^{\top}\big\|^{2}}$ \\
$\sigma_{\min}$ & Minimum required signal strength & 0 \\
\midrule
$\mathcal{A}$ & Action set & \makecell[l]{\{up,right,down,left,hover\}} \\
$R_n$ & Immediate reward at time step n & as Eq.~(1)–(2) \\
$\nu_i$ & Reward weights used in Eq.~(1) & $\sum_{i=1}^{7}\nu_i=1,\ \nu_i\in[0,1]$ \\
\bottomrule
\end{tabularx}
\vspace{-0.4cm}
\end{table}
Finally, we introduce Table~\ref{tab:base_vars_core_strict} to summarize the base variables and notation used in this paper.
\section{Methodology: Improved Noisy Deep Q-Network}
\subsection{Solution via Reinforcement Learning}
In this section we describe our proposed solution for the UAV trajectory navigation problem. As we mentioned before, we solve the problem with our proposed RL method. In RL methods we have four important components: agent, environment, states, and actions. For each step, the UAV state vector $\mathbf{s}_n$ at step $n$ is defined as:
\begin{equation}
    \mathbf{s}_n = \left[ (x_n, y_n)^{T}, \; \left\{ (x_{i}, y_{i})^{T} \right\}_{i \in \mathcal{I}}, \sigma_n \right].
\end{equation}
The action space ${\mathcal{A}}$ consists of five discrete actions:
\begin{equation}
   {\mathcal{A}} = \{a_0, a_1, a_2, a_3, a_4\},
\end{equation}

The state transition for each action is defined as:
\begin{equation}
    \begin{aligned}
        a_0 &: (x_n, y_n) \rightarrow (x_n, y_n+1) \text{ (up)} \\
        a_1 &: (x_n, y_n) \rightarrow (x_n+1, y_n) \text{ (right)} \\
        a_2 &: (x_n, y_n) \rightarrow (x_n, y_n-1) \text{ (down)} \\
        a_3 &: (x_n, y_n) \rightarrow (x_n-1, y_n) \text{ (left)} \\
        a_4 &: (x_n, y_n) \rightarrow (x_n, y_n) \text{ (hover)}.
    \end{aligned}
\end{equation}
The reward function in this study is carefully designed to guide the UAV toward efficient, safe, and communication-aware navigation. At each step, the agent receives a composite reward that integrates multiple factors. The total reward ${R}_n$ is formulated to jointly capture the optimization objective while imposing penalties for constraint violations. It is expressed as follows:
\begin{align}
 R =  & -\nu_{1}D_n  
+ \nu_{2} \left\| (x_n,y_n)^T - (x_{n-1},y_{n-1})^T \right\|^2 \nonumber \\
& + \nu_{3} \left(  T_{\max} - \left\| (x_n,y_n)^T - (x_{n-1},y_{n-1})^T \right\|^2 \right) \nonumber \\
& + \nu_{4}  (x_n,y_n)^T  + \nu_{5} \left( (A-1, A-1)^T - (x_n,y_n)^T \right) \nonumber \\
& + \nu_{6}\sum_{i \in \mathcal{I}}\left(\left\| (x_n,y_n)^T - (x_{i},y_{i})^T \right\|^2 - D_{\min}\right) \nonumber \\
& + \nu_{7}\left( \sigma_n - \sigma_{\min}\right).
\label{reward1}
\end{align}
The final step reward at time step n is given by:
\begin{equation}\label{reward2}
{R}_n=
\begin{cases}
R, & \text{if constraint \text{C2} is satisfied},\\
-|R|, & \text{otherwise}.
\end{cases}
\end{equation}
In this formulation, penalty terms are scaled by weighting factors $\nu_i$, which balance the contributions of different components with heterogeneous units. These weights satisfy $\sum_{i = 1}^{7} \nu_{i} = 1$ and $\nu_{i} \in [0, 1]$.
%
Specifically, the weights are set according to navigation objectives: safety terms are weighted higher, and secondary goals are weighted lower to avoid dominating learning. The final weights are obtained through experimental optimization to ensure the algorithm's stability and reliability.
 This reward structure reflects a balance between exploration and exploitation, where the agent is incentivized to reach the goal efficiently while preserving communication quality and avoiding hazards in a dynamically constrained environment.

We have defined termination conditions for the UAV to improve task completion, prevent infinite loops, and punish behaviors such as collisions. In our UAV navigation setup, the process is terminated when the UAV reaches the target location, exceeds the maximum time limit, or collides with an obstacle (7).
{
\setlength{\abovedisplayskip}{3pt}
\setlength{\belowdisplayskip}{3pt}
\setlength{\abovedisplayshortskip}{2pt}
\setlength{\belowdisplayshortskip}{2pt}
\begin{equation}
    \text{terminal}(\mathbf{s}_n) = \begin{cases}
        1, & \text{if } (x_n, y_n) = (x_{goal}, y_{goal}) \\
        1, & \text{if } n \geq T \\
        1, & \text{if } \text{collision}(\mathbf{s}_n) = 1 \\
        0, & \text{otherwise}.
    \end{cases}
    \label{eq:Terminal}
\end{equation}
}
\subsection{DQN Framework}
RL trains an agent to maximize long-term return through environment interaction. While tabular Q-learning converges in low-dimensional, discrete settings, it suffers from the curse of dimensionality and function-approximation instability in high-dimensional perception and complex dynamics. DQN address these challenges by approximating the action-value function with deep neural networks in a model-free, off-policy paradigm, combining \emph{experience replay} to decorrelate samples and a \emph{target network} to stabilize bootstrapped targets, leading to landmark results in pixel-based discrete control \cite{mnih2015human}.

Formally, the optimal action-value function \(Q^{*}(s,a)\) for a Markov Decision Process (MDP) satisfies the Bellman optimality equation:
\begin{equation}
Q^{*}(s,a) \;=\; \mathbb{E}\!\left[\, r + \gamma \max_{a'} Q^{*}(s',a') \,\middle|\, s,a \right],
\end{equation}
where \(r\) is the immediate reward and \(\gamma\in(0,1)\) the discount factor. DQN parameterizes \(Q_\theta(s,a)\) and learns \(\theta\) by minimizing the temporal-difference (TD) loss over transitions \((s,a,r,s')\) sampled from a replay buffer \(\mathcal D\). Using a frozen target network \(Q_{\theta^-}\), the learning target is
{
\setlength{\abovedisplayskip}{3pt}
\setlength{\belowdisplayskip}{3pt}
\setlength{\abovedisplayshortskip}{2pt}
\setlength{\belowdisplayshortskip}{2pt}
\begin{equation}
y \;=\; r + \gamma \max_{a'} Q_{\theta^-}(s',a'),
\end{equation}
}
and the objective is
\begin{equation}
\mathcal{L}(\theta) \;=\; \mathbb{E}_{(s,a,r,s')\sim \mathcal D}\!\left[(\,y - Q_\theta(s,a)\,)^2\right],
\end{equation}
optionally replaced by the Huber loss for robustness. The target network is updated periodically (hard update) or via soft updates, e.g.,
\begin{equation}
\theta^- \leftarrow \tau \theta + (1-\tau)\theta^-, \quad 0<\tau\ll 1,
\end{equation}
to mitigate target drift. Exploration is typically realized by \(\varepsilon\)-greedy, but its state-agnostic randomness is decoupled from value estimation, which can yield inefficient or homogeneous exploration.

Subsequent work injects learnable, parameterized noise directly into the value network’s parameters (\emph{Noisy DQN}) \cite{fortunato2018noisy}, which improving stability and sample efficiency—and sets the stage for the extensions developed in the next section.
\subsection{Network Architecture with Learnable Noise}
\begin{figure}[htbp]
  \centering
  \includegraphics[scale=0.46]{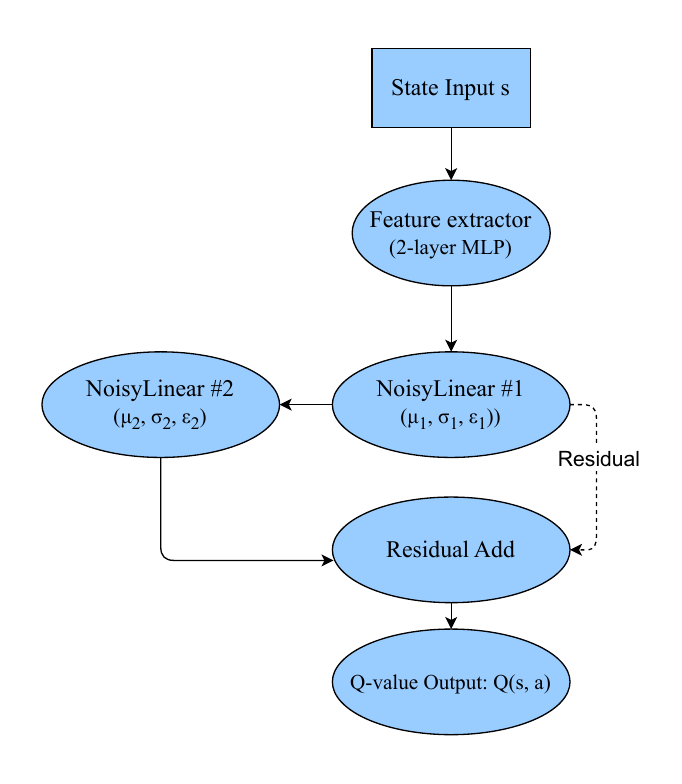}
  \caption{Improved Noisy DQN Network Architecture.}
  \label{fig:structure}
\end{figure}

Building on the noisy network framework, which was first proposed by Fortunato et al.~\cite{fortunato2018noisy}, we propose several improvements. The Q-network $Q(\mathbf{s}, a; \theta, \zeta)$ includes a two-layer feature extractor, two NoisyLinear layers with residual connections, and a value head. The NoisyLinear replaces the standard linear transformation by converting each layer's output into a function with noise, rather than a fixed function. Thus, the result of each forward propagation is different. This introduces the exploratory nature of the strategy during training. The linear transformation with noise is defined as:
\begin{equation}
y = \left( \mu^w + \alpha \cdot \sigma^w \odot \epsilon^w \right) x + \left( \mu^b + \alpha \cdot \sigma^b \odot \epsilon^b \right).
\label{eq:nosiy}
\end{equation}
where $\mu^w, \mu^b$ are learnable means, $\sigma^w, \sigma^b$ are learnable standard deviations, $\epsilon^w, \epsilon^b$ are sampled from standard normal distribution, and $\alpha$ is the noise scale factor. To reduce computational complexity, we adopt factorized Gaussian noise:
\begin{equation}
\epsilon^w_{ij} = f(\epsilon_i) \cdot f(\epsilon_j),
\end{equation}
\begin{equation}
f(x) = \mathrm{sign}(x) \cdot \sqrt{|x|}.
\end{equation}
Our Improved Noisy DQN network architecture is illustrated in Fig.~\ref{fig:structure}.

\vspace{-8pt}
\subsection{Adaptive Noise Scheduling Mechanism}
To regulate exploration over time, we introduce a noise decay schedule with optional performance-based feedback. The noise scaling factor $\alpha(n)$ is defined as:
\begin{equation}
\alpha(n) = \left[ \alpha_{\min} + (\alpha_{\max} - \alpha_{\min}) \cdot \left(1 - \frac{n}{T_{\mathrm{decay}}} \right) \right] \cdot g(P_n),
\end{equation}
where $n$ is the training step, $T_{\mathrm{decay}}$ is the decay duration, and $P_n$ is the recent success rate. The adjustment function $g(P_n)$ is defined as:
\begin{equation}
g(P_n) = 
\begin{cases}
1.2, & P_n < 0.3 \\
0.9, & P_n > 0.8 \\
1.0, & \text{otherwise}.
\end{cases}
\end{equation}
We periodically resample noise in all learnable layers every $k$ steps to promote behavioral diversity during learning.
\begin{equation}
\text{if } n \bmod k = 0, \quad \text{reset\_noise()}.
\end{equation}
\subsection{Target Value Estimation with Double DQN}
To prevent overestimation bias, We use the Double DQN strategy to estimate the target Q-value:
\begin{equation}
y_n = {R}_n + \gamma \cdot Q_{\mathrm{target}}\left( \mathbf{s}_{n+1}, \arg\max_a Q(\mathbf{s}_{n+1}, a; \theta, \zeta) \right),
\end{equation}
where $\gamma \in [0,1]$ is the discount factor, $Q_{\text{target}}$ denotes the target network, and $\theta, \zeta$ are the standard weights and noise parameters of the Q network. The loss is the mean squared error between the current Q-value and the target:
\begin{equation}
\mathcal{L}_n = \left( y_n - Q(\mathbf{s}_n, a_n; \theta, \zeta) \right)^2.
\end{equation}
where $Q(\mathbf{s}_n, a_n; \theta, \zeta)$ is the estimated Q-value for the action $a_n$ taken at state $\mathbf{s}_n$ using the current noisy Q-network.

To stabilize training, a smoothed loss is maintained:
\begin{equation}
\hat{\mathcal{L}}_n = \lambda \cdot \mathcal{L}_n + (1 - \lambda) \cdot \hat{\mathcal{L}}_{n-1}.
\end{equation}
where $\lambda$ is the smoothing coefficient, and $\hat{\mathcal{L}}_{n-1}$ is the previous step’s smoothed loss. The smoothed loss $\hat{\mathcal{L}}_n$ is used to reduce the high variance of instantaneous TD errors during training. This recursion implements an exponential moving average of the loss $\mathcal{L}_n$, which helps stabilize gradient updates and improve overall training robustness.

\vspace{-3pt}
\subsection{Soft Target Network Updates}
We apply a soft update mechanism to gradually align the target network with the main Q-network:
\begin{equation}
\theta_{\mathrm{target}} \leftarrow \tau \cdot \theta + (1 - \tau) \cdot \theta_{\mathrm{target}}.
\end{equation}
where $\tau = 0.005$ is the update rate. This approach reduces the instability caused by the abrupt target shifts.
\subsection{Learning Rate Warm-Up and Cosine Annealing}
In order to optimize the training efficiency and stability, we have introduced a compound learning rate schedule, which consists of a linear warm-up followed by cosine annealing:
\begin{equation}
\eta_n =
\eta_0 \cdot 
\begin{cases}
\frac{n}{T_{\text{warmup}}}, & n \le T_{\text{warmup}} \\
\frac{1 + \cos\left( \pi \cdot \frac{n - T_{\text{warmup}}}{T_{\text{total}} - T_{\text{warmup}}} \right)}{2}, & n > T_{\text{warmup}}.
\end{cases}
\end{equation}
where $\eta_n$ is the learning rate at step $n$, $\eta_0$ is the initial (maximum) learning rate, $T_{\text{warmup}}$ defines the number of warm-up steps, and $T_{\text{total}}$ is the total training steps. 
This schedule avoids the instability during the early training process and enables more precise convergence in the later stages.
\subsection{Training Algorithm}
The full training procedure is summarized in Algorithm~\ref{alg:training}.

\begin{algorithm}[H]
\footnotesize
\linespread{0.95}\selectfont
\caption{Training Procedure of Improved Noisy DQN}
\label{alg:training}
\begin{algorithmic}[1]
\State Initialize Q-network parameters $\theta$
\State Initialize target network $\theta^{-} \gets \theta$
\State Initialize replay buffer $\mathcal{D}$
\State Initialize noise scale $\alpha \gets \alpha_{\max}$

\For{each training step $n = 1$ to $T$}
    \State Observe state $\mathbf{s}_n$
    \State Select action $a_n \gets \arg\max_a Q(\mathbf{s}_n, a; \theta, \zeta)$
    \State Execute $a_n$, observe ${R}_n$, $\mathbf{s}_{n+1}$, and $done$
    \State Store transition $(\mathbf{s}_n, a_n, {R}_n, \mathbf{s}_{n+1}, done)$ in $\mathcal{D}$

    \If{$|\mathcal{D}| >$ batch\_size}
        \State Sample mini-batch from $\mathcal{D}$
        \State Compute target $y_n$ using target network
        \State Compute loss $L$ between $Q(\mathbf{s}_n, a_n)$ and $y_n$
        \State Apply loss smoothing
        \State Backpropagate and update $\theta$
        \State Adjust learning rate

        \If{$n \bmod k = 0$}
            \State Reset noise in all noisy layers
        \EndIf

        \If{using soft update}
            \State Update target network softly
        \ElsIf{$n \bmod C = 0$}
            \State Replace target network parameters
        \EndIf
    \EndIf

    \State Update noise scale $\alpha(n)$
\EndFor

\end{algorithmic}
\end{algorithm}
\subsection{Performance Metrics}
The reward reflects the overall gain of the agent in each training episode, measuring the overall effectiveness of its decision-making within the environment. Specifically, for episode $n$, the total reward $R_{total}$ is defined as the sum of all immediate rewards received at each step during that episode:
$R_{total} = \sum_{k=0}^{T_n} R_n$, where $R_n$ is the immediate reward at step $n$, and $T_n$ is the total number of steps taken in episode $n$. $T_n$ is also applied during training, measures the number of actions required by the agent to complete the task. If the agent has taken $T_{\text{max}}$ steps, it implies that the task is highly likely to have failed.
\vspace{-4pt}
%
%

\section{Simulation Results}
To evaluate the effectiveness of NoisyNet DQN, we evaluated Improved NoisyNet DQN compared with Standard DQN, Standard NoisyNet DQN, and Double DQN in a simulated 15×15 UAV navigation environment. Each agent was trained for 5000 episodes, and the performance metrics included total reward and the number of steps used per episode.
\begin{figure}[htbp]
  \centering
  \includegraphics[width=0.8\linewidth]{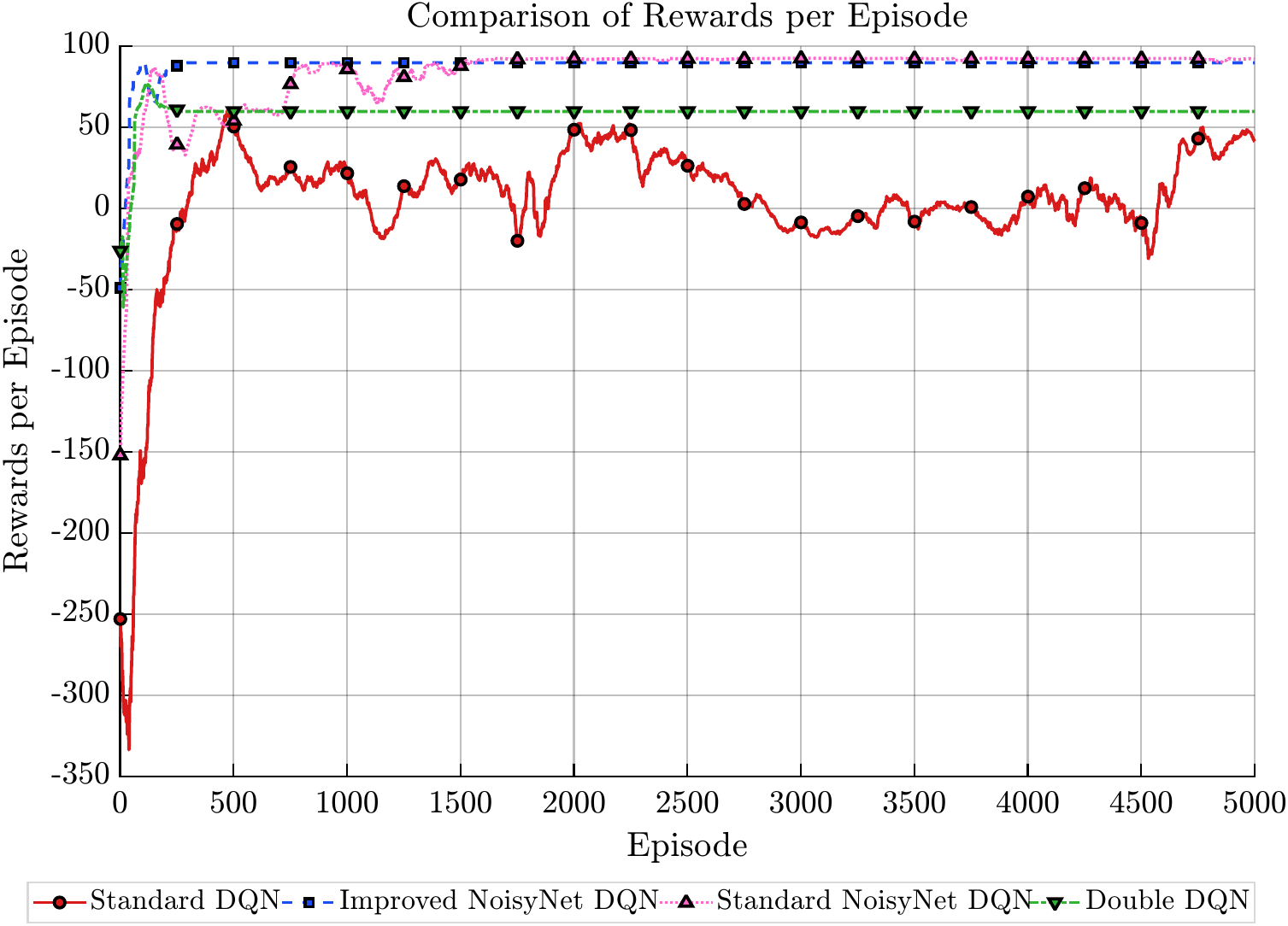}
  \caption{Learning Curve for Reward Comparison Between Noisy DQN and Other Variations.}
  \label{fig:reward_plotfig:steps_plot}
\end{figure}

\begin{figure}[htbp]
  \centering
  \includegraphics[width=0.8\linewidth]{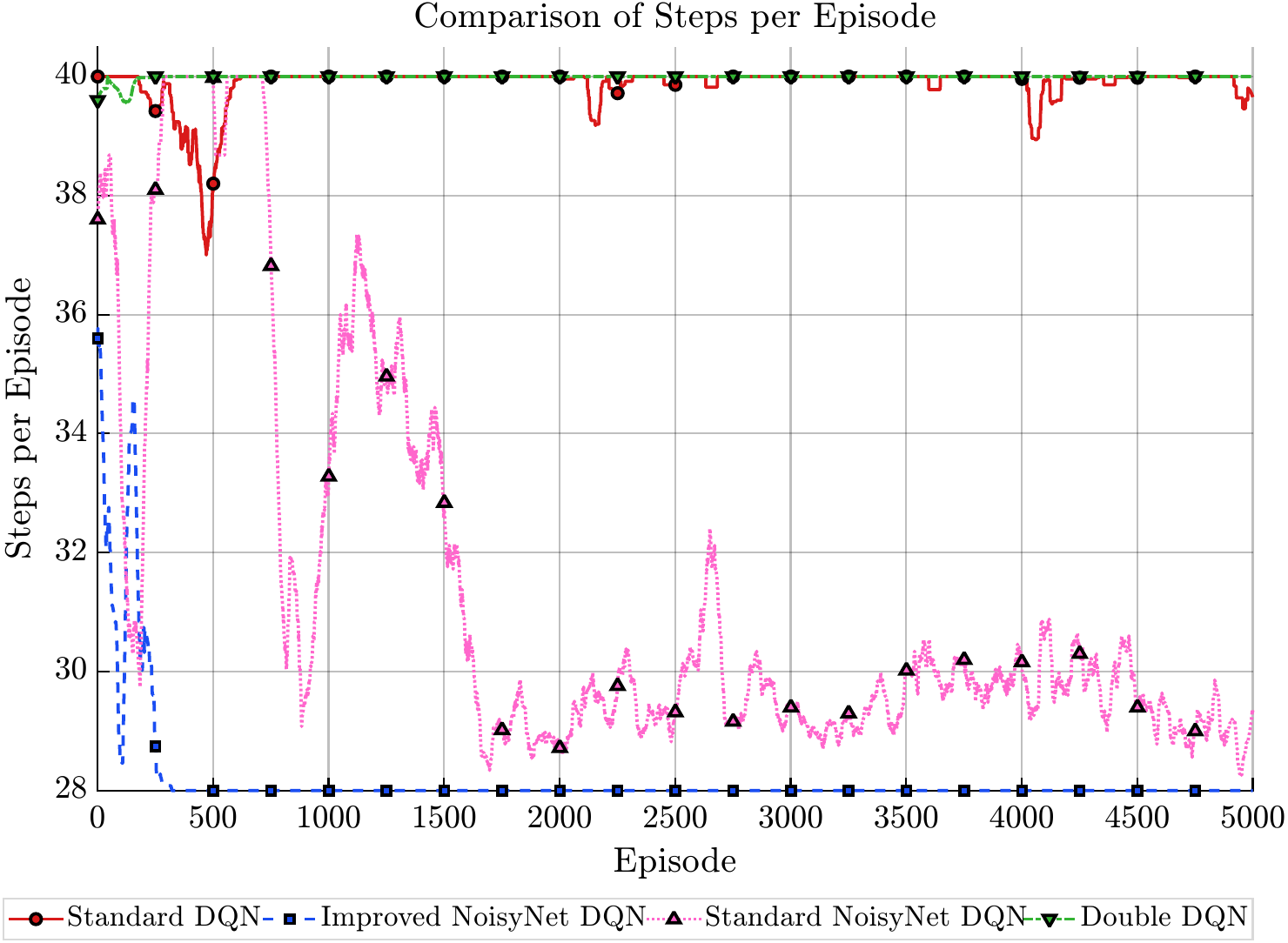}
  \caption{Number of Steps Taken Per Episode.}
  \label{fig:steps_plot}
\end{figure}

The experimental results show that the Improved Noise DQN significantly outperforms the other three DQN algorithms. Fig.~\ref{fig:reward_plotfig:steps_plot} shows that the Improved Noisy DQN quickly reaches a stable high-reward of nearly 100 with the smallest variance; Standard Noisy DQN achieves a similar level of rewards, but it takes more episodes, and is not as stable as the improved algorithm before reaching convergence. Double DQN converges later with moderate variance, and Standard DQN exhibits the largest oscillations and the slowest stabilization, with the final reward value being approximately 60. Fig.~\ref{fig:steps_plot} highlights efficiency: Improved Noisy DQN drives the step count down to the task’s lower bound fastest and maintains it most consistently; Standard NoisyNet DQN drops to a range of 28-30 steps with significant fluctuation. Although Double DQN, DQN and Standard DQN attempt to complete the task through effective steps, most episodes lasted around 40 steps, which indicates that the task was very likely not completed.

Overall, these two figures indicate that injecting parameterized noise into network weights yields more effective exploration and faster, more stable convergence; while Noisy DQN and Double DQN mitigate several shortcomings of DQN, the improved noisy architecture delivers the most consistent gains without sacrificing final performance.

\section{Conclusions}

This paper presents an improved Noisy DQN algorithm for the navigation of UAVs in multi-obstacle environments. This model enhances exploration ability through adaptive noise scheduling, reduces q-value overestimation by combining double DQN, and improves learning stability using the soft update mechanism. The experiments show that the improved NoisyNet DQN achieves faster convergence and fewer steps in task completion than the DQN and other variants. The improved model demonstrates better policy learning efficiency and more stable training dynamics.
However, the framework remains limited by single-step TD learning, making it difficult to fully capture long-term reward signals. This study did not verify performance in the continuous action or high-dimensional state space, limiting its generalization to complex tasks. Future research can explore multi-step value estimation, policy optimization methods based on sequence modeling, and integration with continuous control to enhance the robustness and generalization of the algorithms.

\section{Acknowledgment}
This work was supported by the 2024 Student Partnering with Faculty (SpF) Research Program (Grant WKUSPFS202409), Wenzhou-Kean University.

\begingroup
\footnotesize
\makeatletter
\def\@IEEEbibitemsep{0pt}  
\makeatother
\bibliographystyle{IEEEtran}
\bibliography{Bibliography}
\endgroup

\end{document}